\documentclass[fleqn,twoside]{article}
\usepackage[]{espcrc2}
\usepackage{color}
\definecolor{darkblue}{rgb}{0,0,.5}

\usepackage{hyperref}
\hypersetup{pdftex=true, breaklinks=true,
frenchlinks=true, % Links werden als smallcaps, anstatt farbig dargestellt.
colorlinks=false,
linkcolor=darkblue,% red,% darkblue,
menucolor=darkblue,% red,% darkblue,
pagecolor=darkblue,% red,% darkblue,
urlcolor=darkblue,% red% darkblue
filecolor=darkblue
}
\usepackage[all]{hypcap} % makes figures, tables etc. really visible when pointed to
\usepackage{url} % has to come after \usepackage{hyperref}

\usepackage{graphicx}

\usepackage[latin1]{inputenc}
\usepackage[T1]{fontenc} %\usepackage{txfonts}
\usepackage{microtype} %\usepackage{times}
\usepackage{lmodern} %\usepackage{arial}

\usepackage{amssymb}
\usepackage{mathptmx}
\usepackage{graphicx}
\usepackage{color}
\usepackage{graphicx}
\usepackage{epsfig}
\usepackage[figuresright]{rotating}
\usepackage{amsmath}

\newcommand{\be}{\begin{equation}}
\newcommand{\ee}{\end{equation}}
\newcommand{\bea}{\begin{eqnarray}}
\newcommand{\eea}{\end{eqnarray}}
\newcommand{\eps}{\varepsilon}

\title{%
\vspace*{-2.7cm}
 {\small
 \texttt{DESY 10-077
 \\
 HEPTOOLS 10-019
 \\
 SFB/CPP-10-44\\[1.5cm]
 }}
News on Ambre and CSectors\footnotemark
}
\author{
Janusz~Gluza
\address[KA]{Institute of Physics, University of Silesia, Uniwersytecka 4, PL-40007 Katowice, Poland},
Krzysztof~Kajda
\addressmark[KA],
Tord~Riemann
\address[ZE]{Deutsches Elektronen-Synchrotron, DESY, Platanenallee 6, D-15738 Zeuthen, Germany},
Valery~Yundin
\addressmark[ZE]
}

\runtitle{}
\runauthor{}

\begin{document}

%%%%%%%%%%%%%%%%
\begin{abstract}
Mellin-Barnes and sector decomposition methods are used
to evaluate tensorial Feynman diagrams in the Euclidean
kinematical region.
Few software packages are shortly described and few examples demonstrate their use.
\end{abstract}
%%%%%%%%%%%%%%%
\maketitle \thispagestyle{empty}
%%%%%%%%%%%%%%%%%%%%%%
 \allowdisplaybreaks
\section{Introduction}
%%%%%%%%%%%%%%%%%%%%%%
Mellin-Barnes (MB) methods (see e.g. \cite{Smirnov:2004ym} and references therein) and sector decomposition (SD)
 methods (see e.g. \cite{Heinrich:2008si} and references therein) are frequently used
in theoretical high energy physics calculations.
We will restrict ourselves here mainly to Feynman loop integrals in $d$ dimensions.
There are nowadays public software packages which may help to automatize these calculations.
On the MB Tools webpage \cite{mbtools} there are
useful codes which realize: analytic continuations of Mellin-Barnes integrals -- MB.m
\cite{Czakon:2005rk} and MBresolve.m \cite{Smirnov:2009up};
parametric expansions of Mellin-Barnes integrals -- expansion.m (author M. Czakon) and barnesroutines.m (author D. Kosower).
For the creation of MB representations and later calls of other tools, AMBRE.m can be used
\cite{Gluza:2007rt,ambre:2007}, which we describe here.
AMBRE v1.x generates multiloop scalar [planar]
and one-loop Feynman integrals.
Here we extend the package to v2.0, which may treat also tensor [planar] structures.
For the  sector decomposition method, there are two public programs.
FIESTA \cite{Smirnov:2008py} and FIESTA2 \cite{Smirnov:2009pb}
calculate scalar Feynman integrals with
Mathematica, while Bogner/Weinzierl's (BW) program
\cite{Bogner:2007cr} allows
to calculate basic polynomials needed for the calculation of any Feynman integral
with the c++ based computer algebra system Ginac \cite{Bauer:2000cp1}.
Here we describe the  Mathematica program CSectors.m which may serve as a kind of user interface to BW and Ginac in order to calculate fully automatically tensor Feynman integrals.
%\foot
\renewcommand{\thefootnote}{\fnsymbol{footnote}}
\footnotetext{Talk presented by J.G. at ``Loops and Legs in Quantum Field Theory'', 10th DESY Workshop on Elementary Particle Theory, 25-30 April 2010, W\"{o}rlitz, Germany}
%%%%%%%%%%%%%%%%%%%%%%%%%%%%%%%%%%%%%%%%%%%%%%%%%
\section{General tensor structure \label{sec-tensor}}
%%%%%%%%%%%%%%%%%%%%%%%%%%%%%%%%%%%%%%%%%%%%%%%%%
We consider the $L$ loop Feynman momentum integrals with $N$ propagators $P_n$ and %some
tensor structure $T(k)$:
\begin{eqnarray}
&&G_L[T(k)]\equiv
\frac{1}{(i \pi^{d/2})^{L}}
        \int
        \frac{d^d k_1 \dots d^d k_L T(k)}{P_1 \dots P_N}
\nonumber \\
&=&\frac{1}{(i\pi^{d/2})^{L}} \int \frac{d^dk_1 \ldots d^dk_L~~T(k)}
     {(q_1^2-m_1^2)^{\nu_1} \ldots
       (q_N^2-m_N^2)^{\nu_N}  }
\nonumber \\
&=&  \frac{(-1)^{N_{\nu}}}
  {\Gamma (\nu_{1}) \dots \Gamma (\nu_{N})}
\label{bas1}\nonumber\\
&& \times~  \int
  \prod_{j=1}^{N}dx_j x_{j}^{\nu_{j}-1}
  \delta\left(1-\sum_{i=1}^N x_i\right)
\label{bas2}
\nonumber\\
&&\times~
\sum_{r\le m}
\frac{\Gamma\left(N_{\nu}-\frac{d}{2} L-\frac{r}{2}\right)}{(-2)^{\frac{r}{2}}
}
{{\frac
        {U^{N_{\nu}-\frac{d}{2}(L+1)-m}}
        {F^{N_{\nu}-\frac{d}{2}L-\frac{r}{2}}}
}}\label{bas3}
 \nonumber\\
&& \times~ {{
\left\{{\cal A}_r P^{m-r}\right\}^{[\mu_1,\ldots,\mu_m]}
}}.
\label{bas4}
\end{eqnarray}
The momenta $q_i$ are linear
combinations of internal
momenta $k_j$ and external momenta $p_n$, and $T(k)$ is a tensor structure, e.g.
($1, k_l^{\mu}, k_l^{\mu}k_n^{\nu}, \ldots$).
Further, $N_\nu = \sum \nu_n$.
The general scalar $L$ loop integrals $G_L[1]$  are  implemented in AMBRE.m
v1.x, however tensor structures were allowed only for the one-loop case,  $G_1[T(k)]$.
The symbol $\{R\}^{[a_1\cdots a_r]}$ means the totally symmetrized sum of all possible combinations in the argument.
We give as an explicit example the rank $R=3$ tensor:
        \begin{eqnarray}
&&                \sum_{r\le 3}
                \left\{{\cal A}_r P^{3-r}\right\}^{[\mu_1 \mu_2 \mu_3]}
\\&=&
\left\{{{A_0}} P^3+{{A_1 P^2}}+A_2 P^1+
{{A_3 P^0}} \right\}^{[\mu_1 \mu_2 \mu_3]}
\nonumber \\
&=&P^{\mu_1} P^{\mu_2} P^{\mu_3}+\tilde{g}^{\mu_1 \mu_2}P^{\mu_3}
+\tilde{g}^{\mu_2 \mu_3}P^{\mu_1}
\nonumber \\\nonumber
&& + ~\tilde{g}^{\mu_3 \mu_1}P^{\mu_2} .
 \end{eqnarray}
It is $A_0 = P^0 =1$, and
$A_r$ is zero for
$r$ odd, while ${\cal A}_r= \tilde{g}^{\left[ \mu_1 \mu_2\right.}\cdots \tilde{g}^{\left. \mu_{r-1} \mu_r\right]}$ for $r$ even.
Further,
 $\tilde{g}^{\mu_i \mu_j}=(\tilde{M}^{-1})_{a b}g^{\mu_i \mu_j}$
and
$P_a^{\mu_i}= \sum_l [\tilde{M}_{al}Q_l]^{\mu_i}$.
The $\tilde{M}$ and $Q_l$ are defined from
$\sum_{i=1}^{n} P_i x_i =
        \sum_{i=1}^{N} (q_i^2 - m_i^2)x_i
        =
        \sum_{i,j=1}^L k^T_i M_{ij} k_j - 2 \sum_{j=1}^L k^T_j Q_j + J$
and $\tilde{M}=\text{det}(M)M^{-1}$.
The $U$ and $F$ polynomials were defined e.g. in \cite{Gluza:2007rt,ambre:2007}.

%%%%%%%%%%%%%%%%%%%%%%%%%%%%%%%%%%%%%%%%%%%%%%%%%%%
\section{The package AMBRE.m}
%%%%%%%%%%%%%%%%%%%%%%%%%%%%%%%%%%%%%%%%%%%%%%%%%%%%%%
In \cite{KGRY:2010} it is shown how the tensor structures introduced
in section~\ref{sec-tensor} lead, in the loop-by-loop approach, to a list of Mellin-Barnes
representations.
For that purpose, the following basic function of the AMBRE.m package
is invoked: \footnote{Kinematics in the form of a list
must be also properly defined.}

\bigskip

\noindent
{\bf MBrepr[\{numerator\},\{propagators\},
\newline
\{internal momenta\}]}.
%  \newline

\bigskip

\begin{figure}[t]
\vspace*{-1cm}
\scalebox{0.25}{\includegraphics{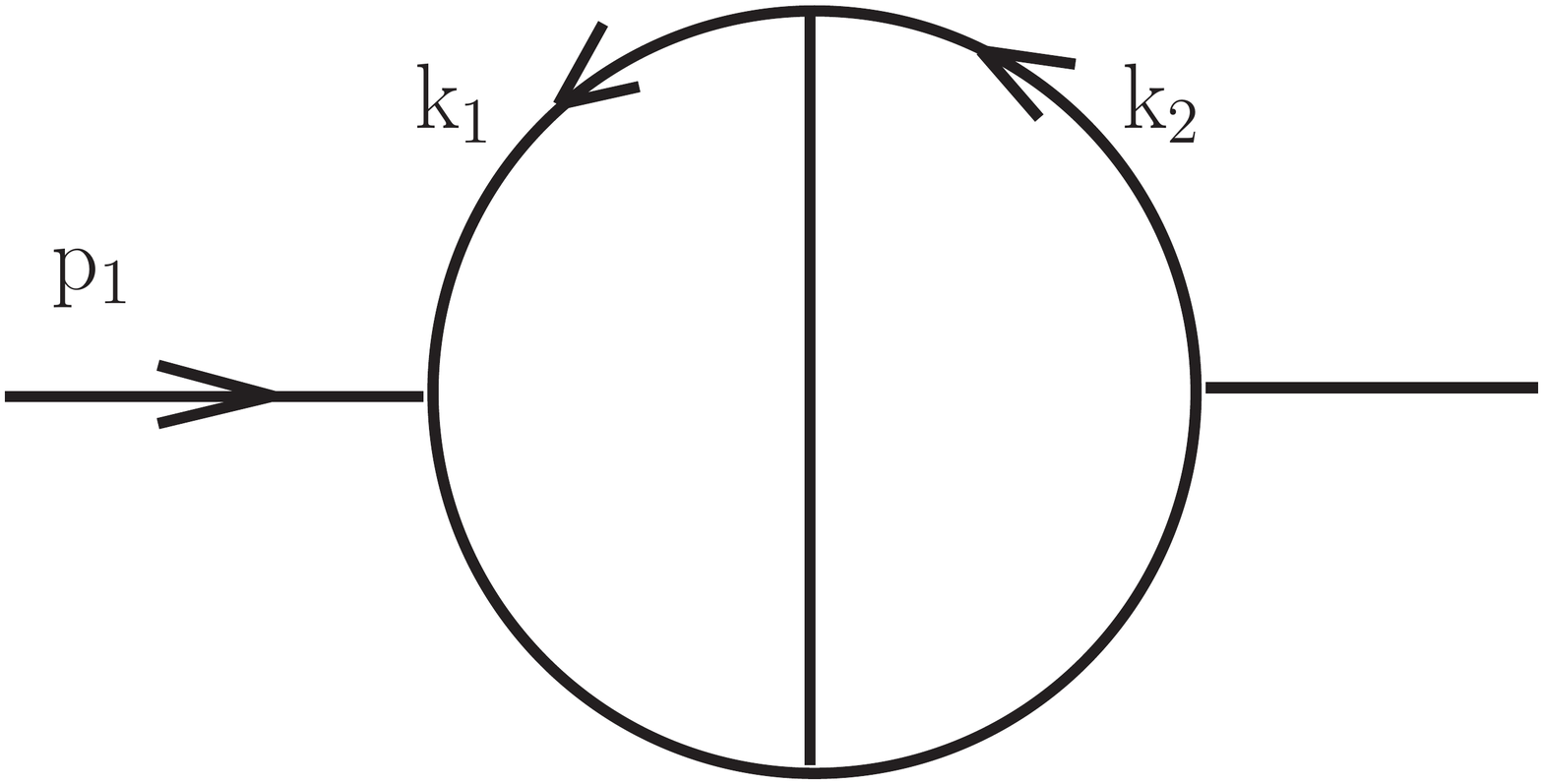}}
\vspace*{-1.5cm}
\caption{A 2-loop self energy topology.}
\label{SEex}
\end{figure}

Let us take the example of figure~\ref{SEex},
and sketch a sample math file.
\footnote{See also the complete MB\_SE5l0m.m source at \cite{ambre:2007}.}

First, packages must be loaded:
\begin{eqnarray}
&&\texttt{<< MBv1.2.m}
\nonumber
\\
&&\texttt{<< AMBRE.m}
\nonumber
\\
&&\texttt{<< MBnum.m}
\label{mbnum}
\end{eqnarray}
or
\begin{eqnarray}
&&\texttt{<< MBresolve.m}
 \label{mbres}
\\ \nonumber
&&\texttt{<< barnesroutines.m}
\end{eqnarray}
As indicated, \verb+ barnesroutines.m+ works with version 1.2 of MB.m.
If we want to use this package,  the MB integrals up to the needed power in
$\epsilon$
must be prepared first (by analytic continuation), and the second part of the
sample MB\_SE5l0m.m file may look as follows:
\begin{verbatim}
invariants = {p1^2 -> s};
repr = MBrepr[{k1*p1, k1*p1, k2*p1},
 {PR[k1,0,n1]*PR[k2,0,n2]*PR[k2-k1,0,n3]
 *PR[k1+p1,0,n4]*PR[k2+p1,0,n5]},
  {k2,k1}];
\end{verbatim}
in the case of the MBnum.m package, followed by:
\begin{verbatim}
SetOptions[MBnum, Analytical -> True,
 Numerical -> False, ShowMBrep -> True];
MBanalytic = MBnum[repr, 1, {s -> -11},
 {n1->1, n2->1, n3->1, n4->1, n5->1}, 2];
\end{verbatim}
\begin{eqnarray}
\label{mbnuma}
\end{eqnarray}

or, in the case of the MBresolve.m package, followed by:
\begin{verbatim}
step1=MBresolve[#/.powers,eps]&/@repr;
step2=MBexpand[step1,
       Exp[2*eps*EulerGamma],{eps,0,1}];
MBanalytic=MBmerge[step2];
\end{verbatim}
\begin{equation}
\label{mbresa}
\end{equation}
Finally, after getting the MB integrals at the $\epsilon$ level needed (here
$\epsilon^1$, second argument of the MBnum function), we ``Simplify'' the
result using Barnes lemmas with the internal MB.m functions MBmerge and MBintegrate:
\begin{verbatim}
after =
Process[MBanalytic, Range[10]]//MBmerge;
SEnumMB=MBintegrate[after,{s->-11}]
\end{verbatim}
The phrase \verb+Range[10]+ means that we assume to get up to 10 dimensional
MB representations; in fact, they are maximally 2-dimensional here. \footnote{Actually
the use of barnesroutines.m does not help much here, and it takes most of the
running time: the time of the total calculation is 17 seconds without applying Barnes Lemmas.}

The final result is:
\begin{eqnarray}
{\rm SEnumMB}
=
-7.5625/\epsilon^2-20.4506/\epsilon
\nonumber \\
-178.18\pm0.0171936
\nonumber \\
+(18.3642\pm 0.0248465) \epsilon .
\label{SEnumMB}
\end{eqnarray}
Singularities are here at most 1-dimensional integrals, so MB.m solves them within
Mathematica very accurately and no error is given.

%%%%%%%%%%%%%%%%%%%%%%%%%%%%%%%%%%%%%%%%%%%%%%%%%%%%%%
\section{The package CSectors.m}
%%%%%%%%%%%%%%%%%%%%%%%%%%%%%%%%%%%%%%%%%%%%%%%%%%%%%%%
The package CSectors.m allows to derive integral representations for Feynman integrals, which are decomposed into so-called sectors with a well-defined singularity structure.
The invariants should be of the Euclidean type.
The structure of the main function is analogous to that in Ambre.m:

\bigskip

\noindent
{\bf DoSectors[\{numerator\},\{propagators\},\newline
\{internal momenta\}][low,des]}

\bigskip

The sample executable file might look as follows: \footnote{See file SD\_SE5l0m.sh at \cite{csectors:2010}.}
\begin{verbatim}
#!/bin/bash
math << \FILE

<< CSectors.m

invariants = {p1^2->s}

SetOptions[DoSectors, SetStrategy -> C,
TextDisplay->True, SourceName->SE5l0m_num,
TempFileDelete->False];

max=0; s=-11;
invariants = {p1^2 -> s};

SEnumSD = DoSectors[{k1*p1,k1*p1,k2*p1},
 {PR[k1,0,1]*PR[k2,0,1]*PR[k2-k1,0,1]*
  PR[k1+p1,0,1]*PR[k2+p1,0,1]},{k2,k1}]
  [-2,max]

FILE
\end{verbatim}
The numerical result in the example is (see also file SD\_LL2010.out at \cite{csectors:2010}):
\begin{eqnarray}
\label{SEnumSD}
{\rm SEnumSD}
=
-178.1927-7.56258/\eps^2
\nonumber\\
- 20.4505/\eps
+
18.394000000000005 \eps,
\nonumber\\
\{0.011130644628528934, 0.00029563/\eps^2,
\nonumber\\
0.00302109/\eps, 0.06629380324170578*\eps\} .
\end{eqnarray}
The second bracket gives an error estimation.
Due to the tensor structure (\ref{bas4}), a number $A$ of
integrals $I_i$ appear at a given $\epsilon$ level.
An error for the numerical result $I_\epsilon$ at $\epsilon$ level
is estimated in a standard way:
\begin{equation}
 \Delta I_\epsilon = \sqrt{ \sum_{i=1}^{A} (\Delta I_i)^2}.
\end{equation}
In the MB.m package the error $\Delta I$ is, for the single MB integral
$I$,  controlled by switching on an appropriate MB.m option:
\verb+Debug->True+.
Let us just note that the error can be underestimated, see \cite{Czakon:2005rk} and MB.m for details.
%%%%%%%%%%%%%%%%%%%%%%%%%%%%%%%%%%
\section{Mellin-Barnes approach versus Sector Decomposition approach}
%%%%%%%%%%%%%%%%%%%%%%%%%%%%%%%%%%%%%
All the results given in the last section in
(\ref{SEnumMB}) and (\ref{SEnumSD})
have been obtained with default parameters needed for numeric defined by
the WB and MB.m packages.
The  total time of calculation on a Xeon PC with 32 GB RAM was  in the MB case
17s
and about 5 minutes for both the C- and X-strategies in the SD case.

\begin{figure}[t]
\begin{center}
\vspace*{-1cm}
\scalebox{0.25}{\includegraphics{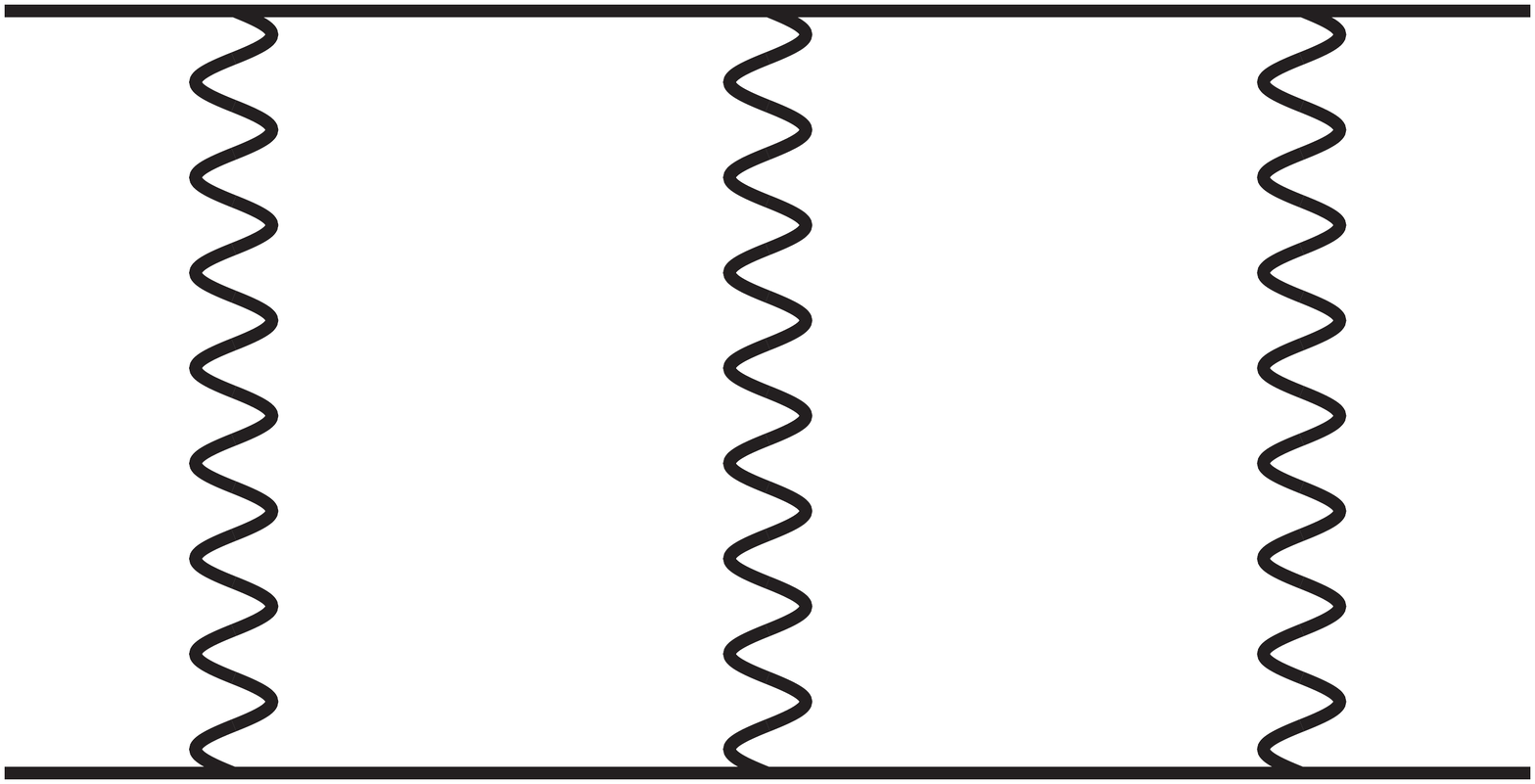}}
\end{center}
\vspace*{-1.5cm}
\caption{The  planar 2-loop Bhabha topology B1.}
\label{B1fig}
\end{figure}

Calculational times can differ substantially depending on methods
and integrals.
In table~\ref{B1massless} and table~\ref{tab-massive},
results for the massless and massive planar 2-loop topology B1
are given, see Fig.~\ref{B1fig} and \cite{Czakon:2004tg,Czakon:2004wm}
for the topology's definition.

\begin{table}[!htb]
   \caption{The scalar massless 2-loop planar double box $B1$, shown in figure \ref{B1fig}.
The scalar case allows for direct comparisons of the packages discussed here with FIESTA2.
For source and output files see \cite{ambre:2007} and \mbox{\cite{csectors:2010}}.\label{B1massless}}
\centering
\begin{tabular}{|r|r|}\hline
massless        & {\texttt{AMBRE}} and {\texttt{MB}} \\
\hline
        $\epsilon^{0}$ & $-0.1034 \pm 6 \cdot 10^{-6}$ \\
        $\epsilon^{-1}$ & $-0.10907$    \\
        $\epsilon^{-2}$ & $-0.00966$    \\
        $\epsilon^{-3}$ & $0.083188$  \\
        $\epsilon^{-4}$ & $-0.022857$   \\
\hline
        T \texttt{[s]} & $22$ \\
\hline
massless      & {\texttt{CSectors, X-strat.}}\\
\hline
        $\epsilon^{0}$  & $-0.1035 \pm 0.0002$  \\
        $\epsilon^{-1}$ & $-0.10915 \pm 0.00008$  \\
        $\epsilon^{-2}$ & $-0.00966 \pm 0.00001$  \\
        $\epsilon^{-3}$ & $0.083191 \pm 0.000005$ \\
        $\epsilon^{-4}$ & $-0.0228574 \pm 1 \cdot 10^{-6}$ \\
\hline
        T \texttt{[s]} & $1712$\\
\hline
massless      & {\texttt{FIESTA2}}\\
\hline
        $\epsilon^{0}$  &  $0.103267 \pm 0.001117$ \\
        $\epsilon^{-1}$ &  $0.109049 \pm 0.000348 $ \\
        $\epsilon^{-2}$ &  $0.009656 \pm 0.000082$\\
        $\epsilon^{-3}$ &  $-0.08319 \pm  0.000012$ \\
        $\epsilon^{-4}$ &  0.022858 \\
\hline
        T \texttt{[s]} & $350$\\
\hline
        \multicolumn{2}{|c|}{$s=-5$, $t=-7$}\\
\hline
\end{tabular}
\end{table}

\begin{table}[!htb]
  \caption{The scalar massive 2-loop planar double box $B1$.
\label{tab-massive}}
\centering
\begin{tabular}{|r|r|}\hline
massive  & {\texttt{AMBRE}} and {\texttt{MB}}  \\
\hline
        $\epsilon^{0}$ & $0.2246$    \\
        $\epsilon^{-1}$ & $0.06359$    \\
        $\epsilon^{-2}$ & $-0.023524$  \\
\hline
        T \texttt{[s]} & $42$ \\
\hline
massive  & {\texttt{CSectors, C-strat.}}\\
\hline
        $\epsilon^{0}$  & $0.2246 \pm 0.0001$  \\
        $\epsilon^{-1}$   & $0.06357 \pm 0.00003$  \\
        $\epsilon^{-2}$   & $-0.023524 \pm 4\cdot 10^{-6}$  \\
\hline
        T \texttt{[s]} & $362$ \\
\hline
massive  & {\texttt{FIESTA2}} \\
\hline
        $\epsilon^{0}$ & $-0.224756 \pm 0.000485$ \\
        $\epsilon^{-1}$ & $-0.063622 \pm 0.000102$ \\
        $\epsilon^{-2}$ & $0.023528 \pm 0.00001$ \\
\hline
        T \texttt{[s]} & $35$ \\
\hline
        \multicolumn{2}{|c|}{$s=-5$, $t=-7$, $m=1$}\\
\hline
\end{tabular}
\end{table}

We can see that in the case of the massless box B1, the MB method can be much faster
than the SD method.
In the massive case the difference is smaller, moreover,
the SD method can be often faster.
In addition results obtained by FIESTA2 \cite{Smirnov:2009pb}
are presented.
Note a difference in sign between FIESTA2 and MB/SD, see \cite{Smirnov:2009pb}
for FIESTA2 conventions.

For completeness, the FIESTA2 sample Mathematica file for getting the  numbers in
Table~\ref{tab-massive} is reproduced:
\begin{verbatim}
<< FIESTA_2.0.0.m;
CIntegratePath="./CIntegrateMP"
QLinkPath = "./QLink64"

invariants =
{p1^2->m^2,p2^2->m^2,p3^2->m^2,p4^2->m^2,
 p1*p2->1/2*s-m^2,p3*p4->1/2*s-m^2,
 p1*p3->1/2*t-m^2,p2*p4->1/2*t-m^2,
 p2*p3->1/2*u-m^2,p1*p4->1/2*u-m^2}
 /.u->4*m^2-s-t/.m->1//Simplify;
kinematics = {s -> -5, t -> -7};

integral=PR[k1,1,n1]*PR[k1+p1,0,n2]*
 PR[k1+p1+p2,1,n3]*PR[k1-k2,0,n4]*
 PR[k2,1,n5]*PR[k2+p1+p2,1,n6]*
 PR[k2 + p1 + p2 + p4, 0, n7];
fintegral=(List @@ integral)
  /.PR[q_,m_,n_]->-(q^2-m^2);

SDEvaluate[UF[{k1,k2},fintegral,
        Join[invariants, kinematics]],
         {1,1,1,1,1,1,1},0]
\end{verbatim}
%%%%%%%%%%%%%%%%%%%%%%%%%%%%%%%%%%%%%%%%%%%%%%%%%%%%%%%%%%%%%%%%%%%%%%%%%%%%%%%%%%%%
\section{Summary}
The CSectors.m package has been prepared in order to allow an easy use of
the SD package by Bogner/Weinzierl for the evaluation of Feynman tensor integrals.
The AMBRE package has been enlarged to involve tensor structures for multiloop
integrals in an automatic way.
We are planning for the near future to
automatize the construction of MB representations for nonplanar diagrams,
what deserves not to use the loop-by loop approach.
Further, we will foresee to build  MB representations for special forms of linear propagators
which are present e.g. in the HQET theory or in calculations of the
QCD static potential (see the talk by V. Smirnov at this conference \cite{Smirnov:2010hn}).

\section*{Acknowledgements}
Work supported in part by Sonderforschungs\-be\-reich/Transregio 9--03 of DFG
``Computergest{\"u}tzte Theo\-re\-ti\-sche Teil\-chen\-phy\-sik'',
Polish Ministry of Science and High Education
from budget for science for years 2010-2013: grant number N N202 102638,
and by MRTN-CT-2006-035505 ``HEPTOOLS'' and MRTN-CT-2006-035482
``FLAVIAnet''.

\end{document}